\documentclass[11pt]{article}

\usepackage{cite}
\usepackage{amsfonts,amsmath,amssymb,mathrsfs,amsthm}
\usepackage{epsfig,epstopdf,graphicx,graphics}
\usepackage{color}
\usepackage{float}
\usepackage{ifpdf}
\usepackage[colorlinks=true]{hyperref}
\usepackage{braket}

\newtheorem{theorem}{Theorem}[section]
\newtheorem{corollary}{Corollary}[section]

\newtheorem{proposition}{Proposition}[section]
\newtheorem{remark}{Remark}[section]

\newcommand{\beq}{\begin{equation}}
\newcommand{\eeq}{\end{equation}}
\newcommand{\beqm}{\begin{equation*}}
\newcommand{\eeqm}{\end{equation*}}
\newcommand{\beqn}{\begin{eqnarray}}
\newcommand{\eeqn}{\end{eqnarray}}
\newcommand{\beqnm}{\begin{eqnarray*}}
\newcommand{\eeqnm}{\end{eqnarray*}}
\newcommand{\bea}{\begin{align}}
\newcommand{\eea}{\end{align}}
\newcommand{\beam}{\begin{align*}}
\newcommand{\eeam}{\end{align*}}
\newcommand{\bs}{\begin{subequations}}
\newcommand{\es}{\end{subequations}}

\newcommand{\bei}{\begin{itemize}}
\newcommand{\eei}{\end{itemize}}
\newcommand{\bed}{\begin{description}}
\newcommand{\eed}{\end{description}}
\newcommand{\bee}{\begin{enumerate}}
\newcommand{\eee}{\end{enumerate}}

\newcommand{\bey}{\begin{array}}
\newcommand{\eey}{\end{array}}

\begin{document}

\title{On the Control of Flying Qubits\thanks{The material in this paper was partially presented at the 20th IFAC World Congress, July 13-17, 2020, Berlin, Germany. RBW acknowledges supports by National Key
Research and Development Program of China (Grant
No. 2017YFA0304300) and NSFC grants (Nos.~61833010 and 61773232). GFZ acknowledges supports in part by Hong Kong Research Grant council (RGC) grants (No. 15208418, No. 15203619) and Shenzhen Fundamental Research Fund, China under Grant No.~JCYJ20190813165207290.
}
}       

\author{Wen-Long Li\thanks{Department of Automation, Tsinghua University, Beijing, 100084, China} \and Guofeng Zhang\thanks{Department of Applied Mathematics, The Hong Kong Polytechnic University, Hong Kong} \and Re-Bing Wu\thanks{Department of Automation, Tsinghua University, Beijing, 100084, China, and Beijing National Research Center for Information Science and Technology, Beijing, 100084, China}}

\date{\today}

\maketitle

\begin{abstract}                          
The control of flying quantum bits (qubits) carried by traveling quantum fields is crucial for coherent information transmission in quantum networks. In this paper, we develop a general framework for modeling the generation, catching and transformation processes of flying qubits.
We introduce the quantum stochastic differential equation (QSDE) to describe the flying-qubit input-output relations actuated by a standing quantum system. Under the continuous time-ordered photon-number basis, the infinite-dimensional QSDE is reduced to a low-dimensional deterministic non-unitary differential equation for the state evolution of the standing system, and the outgoing flying-qubit states can be calculated via randomly occurring quantum jumps. This makes it possible, as demonstrated by examples of flying-qubit generation and transformation, to analyze general cases when the number of excitations is not reserved. The proposed framework lays the foundation for the design of flying-qubit control systems from a control theoretic point of view, within which advanced control techniques can be incorporated for practical applications. 
\end{abstract}

\section{Introduction}
Quantum computation has been evidently demonstrated to be surpassing classical supercomputers in the near future~\cite{Preskill2018}. Towards large-scale and distributed quantum information processing (QIP)~\cite{Duan2001}, the efficient control of networked quantum systems is highly demanding \cite{DiVincenzo1995}. In the past decades, the control of ``standing" components (namely the nodes, e.g., atoms or resonators~\cite{Gu2017}) for on-site QIP had been extensively studied \cite{DAlessandro2008,Dong2010,Wiseman2010,Jacobs2014,Glaser2015,Stefanatos2020}. However, the control of ``flying" components for information transmission, as are often called flying qubits, has received much less attention \cite{Lucamarini2011}.

\begin{figure}
	\centering
	\includegraphics[width=0.9\columnwidth]{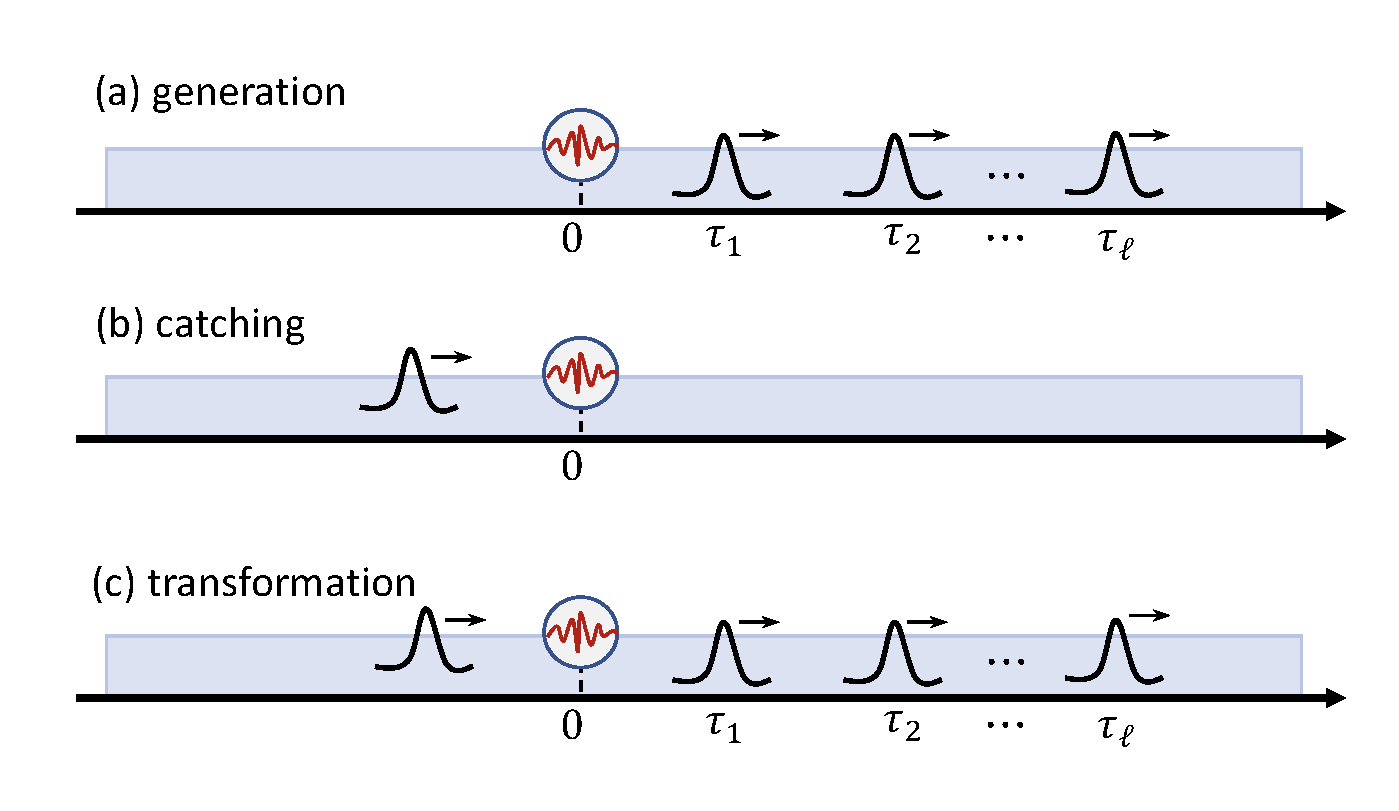}
	\caption{Schematics of flying-qubit control problems actuated by a standing quantum system: (a) the generation of flying qubits; (b) the catching of flying qubits; (c) the transformation of flying qubits.} \label{fig:1}
\end{figure}

In quantum networks, the flying qubits are encoded by the number of photons or the poarization of single photons in travelling electromagnetic fields~\cite{Gheri1998,Cirac1997,Duan2001} ranging from optical regime (e.g., from quantum dots or NV-centers \cite{Kurtsiefer2002,Kuhn2002,Eisaman2011}) down to the microwave regime (e.g., from superconducting artificial atoms~\cite{Houck2007}). Roughly speaking, the flying-qubit control problems can be categorized into the following three classes (see Fig.~\ref{fig:1} for illustration).

The first class is the {\it generation} of flying qubits for the purpose of coding and sending quantum information, the generated flying qubits are required to be on-demand (in the sense that single photons can be triggered at any appointed time) and be flexibly tunable in the frequency~\cite{Peng2016} or the photon shapes~\cite{Keller2004,Pechal2013,Forn-Diaz2017,Averchenko2017,Averchenko2020} so as to match remote network nodes~\cite{Pechal2013,Averchenko2017,Forn-Diaz2017}). 
The second class is the {\it catching} of incoming flying qubits for the purpose of receiving (and storing) quantum information, which is usually done by exciting the standing system from its ground state~\cite{Stobinska2009,Pan2015}. 
The third class is the {\it transformation} of flying qubits (e.g., manipulating the photon number of the shape of incoming flying qubits) for general purposes of information processing~\cite{Srivathsan2014,Leong2016}.

The control of flying qubits need to be actuated by a standing quantum system (e.g., an atom or a resonator~\cite{Yao2005}) of which the incoming and outgoing flying qubits form quantum inputs or outputs. In the generation of flying qubits, the standing system receives vaccum quantum input and flying qubits are yielded as its quantum output. As an inverse process of the generation, the standing system for catching flying qubits receives flying qubits as its quantum input and yields vacuum quantum output. The transformation process, which is also called scattering processes in quantum physics~\cite{Trivedi2018}, can be taken as a generalization of the generation and catching processes, in which both the quantum input and output are non-empty.

To control the flying-qubit input-output processes, one also need classical controls, i.e., tunable time-dependent parameters with some classical devices, e.g., electromagnetic fields that coherently drive the standing system~\cite{Pechal2013,Fischer2018} and tunable couplers that incoherently vary the interaction of the standing system to the waveguide~\cite{Pierre2014}. The control can also be indirectly done by nonlocal spectral filtering using ancilla qubits~\cite{Averchenko2020} or coherent quantum feedback loops~\cite{Dong2016,Zhang2020}. To model the underlying control dynamics, the Quantum Stochastic Differential Equation (QSDE)~\cite{Hudson1984,Gardiner1985} is a powerful tool for analyzing and filtering the input-output responses~\cite{Zhang2011,Zhang2014,Gough2014,Song2016,Dong2019,Zhang2020}, {based on which flying-qubit generaters can be designed with linear or nonlinear finite-level systems~\cite{Gough2012,Gough2014,Gough2015}, as well as the catching of flying qubits via two-level systems with tunable waveguide coupling~\cite{Nurdin2016}.}
However, most existing results~\cite{Yao2005,Shen2009} are heavily dependent on the conservation of energy (or the excitation number) that substantially reduces the effective dimensionality of system. This symmetry will be broken in the presence of time-dependent coherent driving controls as they usually lead to infinite excitations.

{In a preliminary version of this paper~\cite{Li2020}, we found that, under an infinite series expansion, the QSDE approach can be extended to flying-qubit generation systems whose number of excitations is not preserved. Here, we will show that this proposed approach can be further applied to catching and transformation problems with non-vacuum quantum inputs, forming a unified framework for the analysis and design of flying-qubit control systems. The overall contributions we made are as follows. First, we introduce the time-ordered photon-number basis to represent flying-qubit states and their joint state with the standing system. The resulting expansion can be taken as a generalization of the representation proposed in Refs.~\cite{Gough2015,Nurdin2016}, but it is applicable to arbitrary states whose number of excitations are not conservative. Second, as summarized in Theorem \ref{Th:main} and Corollary \ref{corollary} in Sec.~\ref{Sec:basic modeling}, we reduce the original infinite-dimensional QSDE to a low-dimensional deterministic non-unitary evolution equation (see Eq.~(\ref{eq:definition of propagator}) in Sec.~\ref{Sec:basic modeling}) associated with the standing system. This equation can be efficiently solved to calculate the state of outgoing flying qubits via random quantum jumps (see Eq.~(\ref{eq:outgoing state}) in Sec.~\ref{Sec:basic modeling}), which clearly interpret the physical generation process of flying qubits. The resulting model is fully consistent with the results obtained by scattering theory in Ref.~\cite{Fischer2018} for flying-qubit generation, but the derivation is much more concise and easier to generalize to complex quantum networks via $(S,L,H)$ formulation~\cite{Gough2009}. This is quite beneficial to quantum control engineering owing to the prevalent use of the state-space models in control system design. Third, we show that the flying-qubit catching and transformation problems can be unified as generation problems via an auxiliary quantum system that virtually produces the incoming flying qubits, thus forming a systematic QSDE-based framework for treating flying-qubit control problems.}

The remainder of this paper is organized as follows. In Sec.~\ref{Sec:Preliminaries}, necessary preliminaries will be provided for the state representation of flying qubits. In Sec.~\ref{Sec:basic modeling}, the model based on a non-unitary evolution equation is derived for calculating the steady-state of the outgoing flying qubits with vacuum or non-vacuum quantum inputs. In Sec.~\ref{Sec:examples}, we demonstrate the model with applications to flying-qubit generation and transformation problems, respectively, with a two-level standing quantum system. Finally, concluding remarks are made in Sec.~\ref{Sec:Conclusion}.

\section{The Quantum State Representation of Flying Qubits}\label{Sec:Preliminaries}

Consider flying qubits carried by single photons in quantized electromagnetic fields traveling in waveguides. The field generally contains continuous bosonic modes whose annihilation and creation operators satisfy the {singular} commutation relations $[b(\omega),b^{\dag}(\omega')]=\delta(\omega-\omega')$, where $\omega\in\mathbb{R}$ is the mode frequency. In practice, we often represent the field with the inverse Fourier transform~\cite{Gardiner1985}
\begin{equation}\label{}
b_\tau=\frac{1}{\sqrt{2\pi}} \int_{-\infty}^{\infty} d\omega\ e^{-i \omega\tau} b(\omega)
\end{equation}
$b(\omega)$, which describes the annihilation of a photon at time $\tau$. The temporal quantum field. $b_\tau$ satisfies the singular commutation relation $[b_\tau,b^{\dag}_{\tau'}]=\delta(\tau-\tau')$, and can be further used to construct the Ito-type quantum Wiener increment
\begin{equation}
dB_\tau=\int_{\tau}^{\tau+d\tau} b_sds,\quad d\tau>0,
\end{equation}
that obeys {Ito's rule}
\begin{equation}\label{eq:Ito's Rule}
\begin{split}
& dB_\tau dB^{\dag}_s=\delta_{\tau,s}d\tau,
\\&dB^{\dag}_\tau dB_s=dB_\tau dB_s=dB^{\dag}_\tau dB^\dag_s=0.
\end{split}
\end{equation}

Using the Ito-type increment $dB^{\dag}_{\tau}$, we can define the state of an arbitrary single flying-qubit as follows:
\beq \label{eq:1-photon}
\ket{1_\xi}_F = \int_{-\infty}^\infty \xi^\tau dB^\dag_\tau\ket{\Omega}_F,
\eeq
where $\ket{\Omega}_F$ is the vacuum state of the quantum field and the time-dependent shape function $\xi^\tau d\tau$ indicates the probability amplitude of observing the photon between $\tau$ and $\tau+d\tau$. Similarly, we can define the $\ell$-photon state~\cite{Fischer2018},
\beq \label{eq:the l-photon wavepacket under the timeordered}
\ket{\ell_\xi}_F =\int_{-\infty}^\infty \int_{-\infty}^{\tau_\ell}\cdots \int_{-\infty}^{\tau_2} \xi^{\tau_1,\ldots,\tau_\ell} d B^\dag_{\tau_1}\cdots d B^\dag_{\tau_\ell} \ket{\Omega}_F,
\eeq
and more generally, any flying-qubit state can be expanded as the superposition of multi-photon states:
\beq \label{eq:flying-qubit states}
\ket{\xi}_F =\sum_{\ell=0}^\infty \alpha_
\ell\ket{\ell_\xi}=\sum_{\ell=0}^\infty\int_{-\infty}^\infty \int_{-\infty}^{\tau_\ell}\cdots \int_{-\infty}^{\tau_2}  \xi^{\tau_1,\ldots,\tau_\ell}d B^\dag_{\tau_1}\cdots d B^\dag_{\tau_\ell} \ket{\Omega}_F,
\eeq
in which $\xi$ is a scalar when $\ell=0$ and the coefficients $\alpha_\ell$ has been absorbed by $\xi^{\tau_1,\cdots,\tau_\ell}$. Here, for simplicity, we abuse the symbolic use of $\xi$ to represent all components in the superposition of multi-photon wavepackets $\{\xi,\xi^{\tau_1},\xi^{\tau_1,\tau_2},...\}$. Under this representation, the probability of observing $\ell$ flying qubits in $\ket{\xi}$ can be calculated by the integral
\begin{equation}\label{eq:P_l}
P_{\ell}=\int_{-\infty}^{\infty}\int_{-\infty}^{\tau_\ell}\cdots\int_{-\infty}^{\tau_2} |\xi^{\tau_1,\cdots,\tau_\ell}|^2 d\tau_{\ell}\cdots d\tau_1.
\end{equation}
The summation $\sum_{\ell=0}^\infty P_\ell$ of these probabilities must be equal to 1, and this can be easily verified by Ito's rule from the normalization condition $_F\langle \xi|\xi\rangle_F=1$.

%
\section{Flying-Qubit Dynamics via Classical Controls}\label{Sec:basic modeling}

Throughout this paper, we always assume that the classical controls on the standing system are imposed from the initial time $t_0=0$, and the standing system is coupled to a chiral waveguide in which all flying qubits propagate rightwards~\cite{Peng2016}. To analyze the time evolution of the joint system, we expand the time-dependent joint state of the system and the flying qubits as follows:
\begin{equation}\label{eq:Psi_t expansion}
\ket{\Psi(t)}=\sum_{\ell=0}^\infty \int_{0}^t\int_{0}^{\tau_\ell} \cdots \int_{0} ^{\tau_2} |\psi^{\tau_1,\cdots,\tau_\ell}(t)\rangle\otimes d B^\dag_{\tau_1}\cdots d B^\dag_{\tau_\ell} \ket{\Omega}_F,
\end{equation}
where $|\psi^{\tau_1,\cdots,\tau_\ell}(t) \rangle$ is the standing system's correlated state with the field when $\ell$ photons are generated at $\tau_1,\cdots,\tau_\ell$. The modified lower and upper bounds of the first integral indicate that, physically, flying qubits can only be generated between time $0$ and the present time $t$. {The expansion contains terms that possess different numbers of excitations, which can be taken as the generalization of the state representation used in Ref.~\cite{Nurdin2016} that involves one and only one excitation.}

In the following, we will introduce the Quantum Stochastic Differential Equation{s} (QSDE{s}) to model the flying-qubit control processes with vacuum or non-vacuum quantum inputs, respectively.

\subsection{Standing system with vacuum quantum inputs}
We start from the simple case that the quantum input of the standing system is empty (e.g., in its vacuum state), i.e., the joint system is initially at $\ket{\Psi(0)} =  \ket{\psi_{0}}\otimes\ket{\Omega}_F$ with $|\psi_0\rangle$ as the initial state of the standing system. Under this circumstance, the evolution of the joint state is governed by the following QSDE of the joint state $|\Psi(t)\rangle$~\cite{Gough2013,Nurdin2016}:
\begin{equation}\label{eq:SQSSE}
d|\Psi (t)\rangle =\left[-iH_{\rm eff}(t) dt+  L(t) dB^{\dag}_t  \right]|\Psi (t)\rangle,
\end{equation}
where $L(t)$ represents the coupling operator of the standing system to the flying qubits, and incoherent classical controls can be realized by varying $L(t)$ in time. The effective {\it non-Hermitian} Hamiltonian
\begin{equation}\label{}
H_{\rm eff}(t) =H(t) -\frac{i}{2} L^{\dagger}(t)L(t),
\end{equation}
where $H(t)$ includes the standing system's internal Hamiltonian and interaction Hamiltonian associated with the driving fields that serve as coherent classical control functions.

Now we present the main Theorem for the analysis of flying-qubit control dynamics.
\begin{theorem}\label{Th:main}
	Let $V(t)$ be the propagator of the differential equation
	\begin{equation}\label{eq:definition of propagator}
	\dot{V}(t) = -i H_{\rm eff}(t) V(t),
	\end{equation}
	with $V(0) = \mathbb{I}$. Then, the correlated states
	\begin{equation}\label{eq:compact form}
	|\psi^{\tau_1,\cdots,\tau_\ell}(t)\rangle = V(t)\tilde L(\tau_{\ell}) \cdots \tilde L(\tau_1)|\psi_{0}\rangle,
	\end{equation}
	for $\ell=0,1,\ldots$, where $\tilde L(\tau_k)=V^{-1}(\tau_k)L(\tau_k)V(\tau_k)$.
\end{theorem}

\emph{Proof.} Differentiate both sides of Eq.~(\ref{eq:Psi_t expansion}), we have
\begin{equation}\label{eq:differential}
\begin{split}
d|\Psi(t) \rangle =&~ \sum_{\ell=0}^\infty  \int_{0}^{t}\int_{0}^{\tau_\ell}\cdots\int_{0}^{\tau_2}| \dot{ \psi}^{\tau_1,\cdots,\tau_{\ell}}(t)\rangle dt \otimes dB^{ \dag} _{\tau_1}\cdots dB^{\dag}_{\tau_{\ell}}\ket{\Omega}_F
\\&+dB^{\dag}_t\sum_{\ell=1}^\infty  \int_{0}^{t}\int_{0}^{\tau_{\ell-1}}\cdots\int_{0}^{\tau_2}|  \psi^{\tau_1,\cdots,\tau_{\ell-1},t}(t)\rangle\otimes dB^{ \dag}_{\tau_1}\cdots dB^{\dag}_{\tau_{\ell-1}}\ket{\Omega}_F,
\end{split}
\end{equation}
where the dot ``$\cdot$" represents the partial derivative with respect to time variable $t$. Replacing Eqs.~(\ref{eq:Psi_t expansion}) and (\ref{eq:differential}) into Eq.~(\ref{eq:SQSSE}), we can obtain
\begin{equation}\label{eq:dynamics of state}
|\dot{\psi}^{\tau_1,\cdots,\tau_{\ell}}(t)\rangle = -iH_{\rm eff}(t) |\psi^{\tau_1,\cdots,\tau_{\ell}}(t)\rangle
\end{equation}
by comparing the terms with $dt$, where $\ell\geq 0$. The comparison of terms with $dB^{\dag}_t$ gives the boundary condition
\begin{equation}\label{eq:boundary condition of state}
|\psi^{\tau_1,\cdots,\tau_{\ell-1},t}(t)\rangle =  L(t)|\psi^{\tau_1,\cdots,\tau_{\ell-1}}(t)\rangle.
\end{equation}
This condition indicates the emission of a photon (or a flying qubit) accompanied with the lowering of the standing system's energy levels by operator $L(t)$, which is also called quantum jumps in open quantum system theory~\cite{Wiseman2010}. The transition of wavepacket functions from $|\psi^{\tau_1,\cdots,\tau_{\ell-1}}(t)\rangle$ to $|\psi^{\tau_1,\cdots,\tau_{\ell-1},t}(t)\rangle$ implies that the number of photons in the field is increased by 1, where the additional photon is emitted at time $t$ from the standing system.

Observing that all vector functions $|\psi^{\tau_1,\cdots,\tau_{\ell}}(t)\rangle $ share the same differential equation (\ref{eq:dynamics of state}) with respect to time $t$, we denote by $V(t)$ their common non-unnitary propagator steered by the effective Hamiltonian $H_{\rm eff}(t)$, as in Eq.~(\ref{eq:definition of propagator}). Consequently, let
\begin{equation} \label{eq: green function}
G(s,s')=V(s)V^{-1}(s')
\end{equation}
be the transition operator between any $s,s'\leq t$. From Eqs.~(\ref{eq:definition of propagator}) and (\ref{eq:dynamics of state}) we have $|\psi(t)\rangle=G(t,0)|\psi_0\rangle$ and
\begin{equation}\label{eq:waveguide-boundary2}
|\psi^{\tau_1,\cdots,\tau_\ell}(t)\rangle  = G(t,\tau) |\psi^{\tau_1,\cdots,\tau_\ell}(\tau)\rangle,
\end{equation}
for any $0 \leq \tau_1\leq \cdots\leq\tau_\ell \leq \tau \leq t $.
Repeatedly using Eq.~(\ref{eq:waveguide-boundary2}) and Eq.~(\ref{eq:boundary condition of state}), we have
\begin{eqnarray}\label{eq:the general solution}
|\psi^{\tau_1,\cdots,\tau_{\ell}}(t)\rangle
&=&
G(t,\tau_{\ell}) |\psi^{\tau_1,\cdots,\tau_\ell}(\tau_{\ell})\rangle
\nonumber
\\
&=& G(t,\tau_{\ell})L(\tau_{\ell}) |\psi^{\tau_1,\cdots,\tau_{\ell-1}}(\tau_{\ell})\rangle\nonumber \\
&\vdots&
\nonumber
\\
&=&
G(t,\tau_{\ell})L(\tau_{\ell}) G(\tau_{\ell},\tau_{\ell-1})L(\tau_{\ell-1})G(\tau_{\ell-1},\tau_{\ell-2})
\cdots  L(\tau_1) G(\tau_1,0)|\psi_{0}\rangle.
\label{eq:july8_iteration}	
\end{eqnarray}
The compact form (\ref{eq:compact form}) is obtained from Eq.~(\ref{eq:july8_iteration}) by Eq.~\eqref{eq: green function}. $\Box$

\begin{remark}
Theorem \ref{Th:main} characterizes quantum jumps that is well-known in the theory of open quantum systems~\cite{Wiseman2010}, along which single photons are sequentially released at random time instants $\tau_1,\ldots, \tau_\ell$. These quantum jumps intervene in the non-unitary evolution (\ref{eq:dynamics of state}) of $|\psi^{\tau_1,\cdots,\tau_{\ell}}(t)\rangle$ by ${L}(\tau_1),\ldots,{L}(\tau_\ell)$, forming stochastic and discontinuous evolution trajectories that can be numerically calculated to analyze and design the flying-qubit generation processes.\end{remark}

Theorem \ref{Th:main} provides the entire solution for the transient process of generating flying qubits, which is reduced to Eq.~(\ref{eq:definition of propagator}) on the space of the standing quantum system. To calculate the steady-state of the outgoing flying qubits, we have the following corollary.

\begin{corollary}\label{corollary}
If the standing system asymptotically decays to its ground state $|0\rangle$, then the outgoing flying-qubit state can be decomposed into the superposition of multi-photon states represented by Eq.~(\ref{eq:flying-qubit states}), where
\begin{equation}\label{eq:outgoing state}
\xi^{\tau_1,\cdots,\tau_\ell} = \langle 0|V(\infty)\tilde L(\tau_{\ell}) \cdots \tilde L(\tau_1)|\psi_{0}\rangle
\end{equation}
for $\ell=0,1,2,\cdots$.
\end{corollary}

\emph{Proof.} The steady-state of the outgoing flying qubits can be obtained from the asymptotic limit $t\rightarrow \infty$ of the joint state $|\Psi(t)\rangle$. When the atom decays to its ground state $|0\rangle$, only ground-state components in all $\ket{\psi^{\tau_1,\cdots,\tau_\ell}(t)}$ may survive. Hence, the asymptotic state of $|\Psi(t)\rangle$ can be decomposed as
\begin{equation}\label{eq:final state}
|\Psi(\infty) \rangle =|0\rangle\otimes \sum_{\ell=0}^\infty  \int_{0}^{\infty}\int_{0}^{\tau_{\ell}}\cdots\int_{0}^{\tau_2}  \xi^{\tau_1,\cdots,\tau_\ell}dB^{\dag}_{\tau_1}\cdots  dB^{\dag}_{\tau_\ell}\ket{\Omega}_F,
\end{equation}
where $\xi^{\tau_1,\cdots,\tau_\ell}={}\langle0|\psi^{\tau_1,\cdots,\tau_\ell}(\infty)\rangle$. The outgoing flying-qubit state can be directly observed from the flying-qubit part of (\ref{eq:outgoing state}).  $\Box$

\begin{remark}
The assumption made in Corollary \ref{corollary}, i.e., the standing system decays to its ground state, always holds when the system is a two-level atom that is constantly coupled to the waveguide and is driven by a finite-duration driving pulse. Under this circumstance, the standing system will spontaneously decay to its ground state after the control pulse is turned off, and eventually is disentangled from the flying qubits. It is possible that, under elaborately designed coherent and incoherent controls, or with multi-level atoms that have non-unique stable states (e.g., a $\Lambda$-type atom), the standing system remains entangled with the flying qubits. This has many intriguing applications in quantum networks, e.g., distributing quantum entanglements between remote nodes in the quantum network.
\end{remark}


\subsection{Standing system with non-vacuum quantum inputs}

\begin{figure}
	\centering
	\includegraphics[width=0.9\columnwidth]{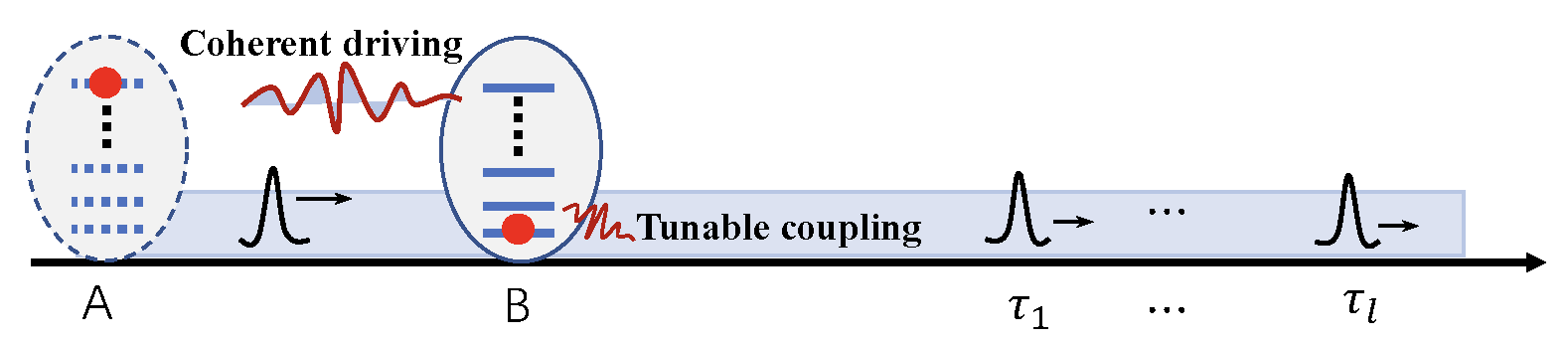}
	\caption{Schematics of a flying-qubit control system $B$ with quantum inputs generated from an auxiliary system $A$.} \label{fig:2}
\end{figure}

The above analysis is well suited for the generation of flying qubits, but it is not directly applicable to the catching or transformation problems because the quantum input of incoming flying qubits is not initially empty. As schematically shown in Fig.~\ref{fig:2}, we can introduce an auxiliary system $A$ that is cascaded to the standing system $B$, where the auxiliary system is engineered to generate a flying qubit at the same state of the incoming flying qubit~\cite{Gough2015}. In this way, the analysis can be equivalently done with the joint $AB$ system with vacuum quantum inputs, to which Theorem \ref{Th:main} and Corollary \ref{corollary} become applicable.

Let $H_A(t)$ and $H_B(t)$ be the Hamiltonians of the $A$-system and $B$-system, and $L_A(t)$ and $L_B(t)$ be their coupling operators to the waveguide, respectively. According to the $(S,L,H)$ formalism presented in \cite{Gough2009}, the series product of systems $A$ and $B$ has the following equivalent Hamiltonian
\begin{equation}\label{eq:equivalent Hamiltonian}
H(t)= H_A(t) \otimes \mathbb{I}_B +  \mathbb{I}_A\otimes H_B(t)   + \frac{1}{2i}\left[L_A(t) \otimes L_B^{\dagger}(t)  -  L_A^{\dagger}(t)\otimes L_B(t)\right],
\end{equation}
and the coupling operator
\begin{equation}\label{eq:coupling operator}
L(t)= L_A(t) \otimes \mathbb{I}_B + \mathbb{I}_A  \otimes  L_B(t).
\end{equation}
The series product representation makes it possible to apply Theorem \ref{Th:main} to analyze flying-qubit catching or transformation problems, because the joint $AB$ system receives vacuum quantum inputs via the subsystem $A$. Note that the auxiliary system $A$ needs to be carefully designed so as to yield the incoming flying qubits to be caught or transformed.

\section{Examples}\label{Sec:examples}
This section will illustrate how the proposed model is applied to the control of flying qubits. For simplicity, the standing system is chosen as a two-level qubit, whose controlled Hamiltonian is
\begin{eqnarray}
H(t) &=& \epsilon(t)\sigma_+\sigma_-+\frac{u(t)}{2} \sigma_++\frac{u^*(t)}{2}\sigma_-, \label{eq:qubit hamiltonian}\\
L(t) &=& \sqrt{2\gamma(t)}\sigma_-, \label{eq:qubit coupling}
\end{eqnarray}
where $\sigma_\pm$ are the standard Pauli raising and lowering operators. The functions $\epsilon(t)$ and $u(t)$ are the detuning frequency and the envelope of the coherent driving field on the qubit. The tunable coupling strength $\gamma(t)$ plays the role of incoherent classical controls. In the following, we discuss the flying-qubit generation and transformation problems, which correspond to the cases with vacuum and non-vacuum quantum inputs, respectively.

\subsection{The generation of flying qubits}
We will first apply Theorem \ref{Th:main} to reproduce a well-known result of generating arbitrary single-photon wavepackets via the incoherent control of tunable waveguide coupling function~\cite{Gough2012,Gough2013,Gough2015}, with a slight generalization in that a time-variant phase is allowed in the wavepacket function.
\begin{proposition}\label{Prop:A}
Given a differentiable single-photon wavepacket $\xi^\tau=\nu(\tau)e^{i\phi(\tau)}$, where $\nu(\tau)$ and $\phi(\tau)$ are the amplitude and phase functions of $\xi^\tau$. Then, a flying qubit at state $\ket{1_\xi}$ can be generated without coherent driving $u(t)$ when the standing qubit is initially prepared at the excited state $|1\rangle$ and
	\begin{equation}\label{eq:the condiany singltion of gene-photon}
	\gamma(\tau) =\frac{\nu^2(\tau)}{2\int_{\tau}^\infty \nu^2(s)ds}, \quad \epsilon(\tau)= -\frac{d\phi(\tau)}{d\tau}.
	\end{equation}
\end{proposition}

\emph{Proof.} According to Theorem \ref{Th:main} and Corollary \ref{corollary}, it is straightforward to calculate the wavepacket of outgoing single-photon state as follows:
\begin{equation}\label{}
\xi^\tau=\sqrt{2\gamma(\tau)}\exp\left[-\int_{0}^\tau\gamma(s)ds \right]\exp\left[-i\int_{0}^\tau\epsilon(s)ds \right].
\end{equation}
Comparing with the shape function $\xi^\tau=\nu(\tau)e^{i\phi(\tau)}$, we have $-\int_0^t\epsilon(\tau)d\tau=\phi(t)$ and
\begin{equation}\label{eq:c0}
\nu(\tau)=
\sqrt{2\gamma(\tau)}\exp\left[-\int_{0}^\tau\gamma(s)ds \right]
\end{equation}
for all $0<\tau<t$. Combined with the normalization condition of $\nu(\tau)$, the amplitude and phase conditions in the proposition can be derived. $\Box$

The proof of Proposition \ref{Prop:A} is relatively simple because the controls $\gamma(t)$ and $\epsilon(t)$ preserve the number of excitations in the joint system. Next, we consider the case with a coherent driving pulse that does not preserve the number of excitations. Assume that the driving control posed along the $x$-axis is resonant with the qubit (i.e., $\epsilon(t)\equiv 0$) and the coupling strength $\gamma(t)\equiv \gamma$. Under a rectangular driving control pulse on $0\leq t\leq T<\infty$, the effective Hamiltonian can be written as
\begin{equation}
H_{\rm eff}(t)=
\begin{cases}
\frac{\Omega}{2} \sigma_x - i\frac{\gamma}{2} \sigma_+\sigma_-, & 0< t < T
\\ - i\frac{\gamma}{2} \sigma_+\sigma_-, & t> T
\end{cases}
\end{equation}
where $\sigma_x=\sigma_++\sigma_-$ and $\Omega$ is the power of the driving pulse.

Depending on the magnitude of the driving power $\Omega$, we discuss the outgoing flying-qubit states in three regimes, namely the strong-driving regime ($\Omega>\gamma$), the balanced-driving regime ($\Omega=\gamma$) and the weak-driving regime ($\Omega<\gamma$). According to Theorem~\ref{Th:main} and Corollary~\ref{corollary}, we can obtain, for the example of strong-driving case, that
\begin{equation}
\xi=\langle0|V(\infty)|0\rangle=e^{-\frac{ \gamma T}{2}}\cos \left(\frac{\omega T}{2}-\phi\right)
\end{equation}
and the single-photon packet
\begin{equation}
\xi^{\tau} =\left\{
\begin{array}{ll}
-\frac{i  \sqrt{ 2\gamma} e^{-\frac{\gamma T}{2}}}{\cos^2\varphi} \sin\frac{\omega {\tau}}{2} \cos\left(\frac{\omega (T-{\tau})}{2} -\varphi\right), &  {\tau}<T \\
-\frac{i\sqrt{ 2\gamma} e^{-\frac{ \gamma T}{2} } }{\cos\varphi} \sin{\frac{\omega T}{2}}e^{- \gamma ({\tau}-T)},
&{\tau} >T,
\end{array}
\right.
\end{equation}
where $\varphi=\arcsin{\frac{\gamma}{\Omega}}$ and $\omega =\sqrt{\Omega^2-\gamma^2}$. The expressions for multi-photon emission, as well as those in balanced-driving and weak-driving regimes, can also be analytically solved under retangular driving pulses, but they will not be listed here due to the limit of space.

\begin{figure}
	\centering
	\includegraphics[width=1\columnwidth]{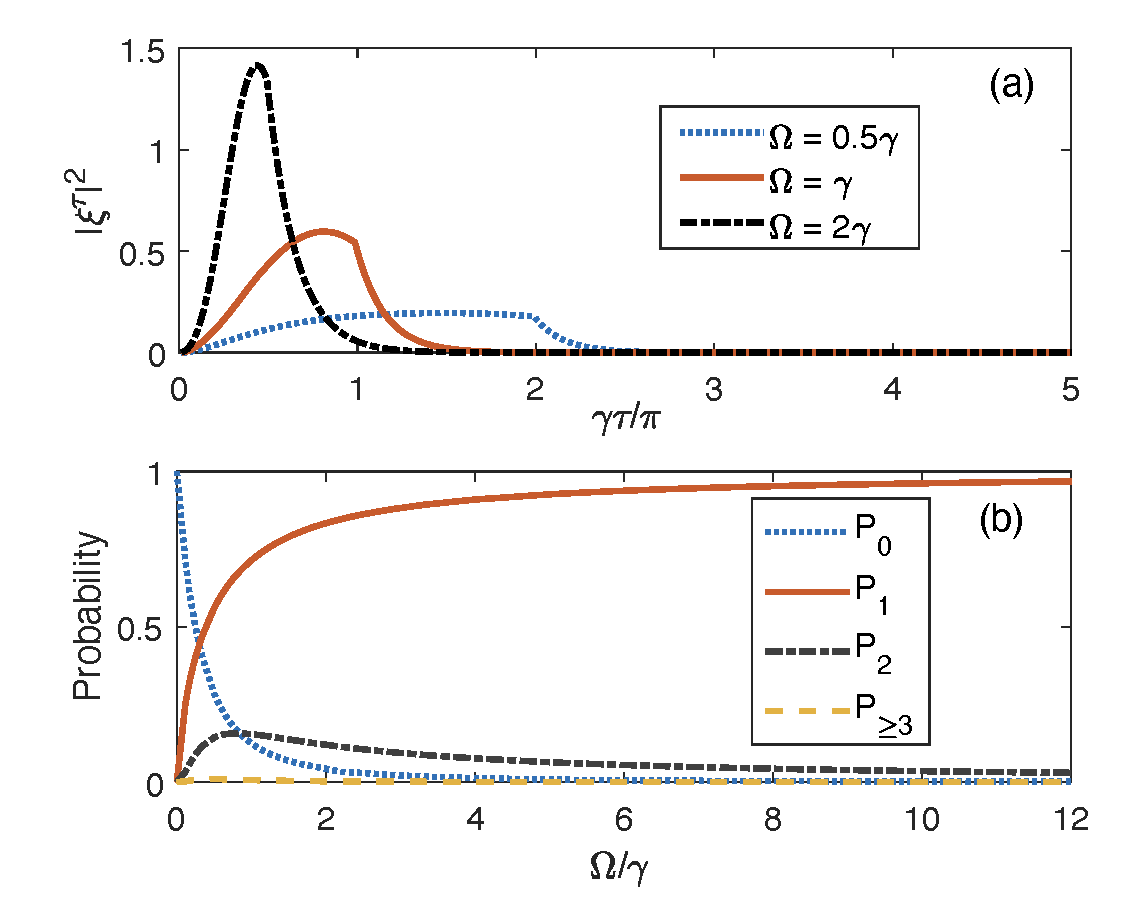}
	\caption{(a) The shapes of single-photon wavepackets generated by $\pi$-pulses in the strong, balanced and weak driving regimes; (b) the probabilities of generating $0$, $1$, $2$ and more flying qubits under $\pi$-pulses with driving strength varying from weak to strong coupling regimes.} \label{fig:3}
\end{figure}

Based on the above calculation, we display in Fig.~\ref{fig:3}(a) the shape $|\xi^\tau|^2$ of the single-photon component generated by soft $\pi$-pulses (i.e., $\Omega T=\pi$) in the strong, balanced and weak driving regimes, respectively. All these wavepackets gradually rise from zero and the steepness of the rising slope is proportional to the driving strength $\Omega$. After the driving pulse is turned off, they spontaneously decay to zero. We also calculate the probabilities of generating $0$, $1$, $2$ and more photons according to Eq.~(\ref{eq:P_l}) when the pulse power $\Omega$ increases from weak- to strong-driving regimes. As is shown in Fig.~\ref{fig:3}(b), soft $\pi$-pulses do not generate perfect single flying qubits due to the coexisting multi-photon emission processes, unless in the hard-pulse limit that $\Omega$ approaches infinity.

\subsection{The transformation of flying qubits}
Now let us see how a standing qubit responses to non-vacuum quantum inputs, e.g., an incoming flying qubit at state $\ket{1_{\xi_0}}$, where $\xi_0^\tau$ is its wavepacket. Due to the limit of space, we only discuss the transformation problems here, as the catching problem can be taken as one of its special case.

Suppose that the standing qubit is prepared at the ground state $|0\rangle$ and is controlled by the tunable coupling $\gamma(t)$ and the coherent driving $u(t)$. We introduce the auxiliary two-lelvel system $A$ shown in Fig~(\ref{fig:2}) to generate the incoming flying qubit, which is initially excited and its Hamiltonian $H_0(t)=\epsilon_0(t)\sigma_+\sigma_-$ and coupling operator $L_0(t)=\sqrt{2\gamma_0(t)}\sigma_-$ are chosen according to Proposition~\ref{Prop:A}. From Eq.~\ref{eq:equivalent Hamiltonian}, the effective Hamiltonian of the $AB$ system is
\begin{equation}
H_{\rm eff}(t)=\frac{u(t)}{2}\sigma_++\frac{u^*(t)}{2}\sigma_-
- i\left[\gamma_A(t) \sigma_+ \sigma_-  \otimes \mathbb{I}_2  + \mathbb{I}_2 \otimes\gamma_B(t) \sigma_+ \sigma_-+2\sqrt{\gamma_0(t)\gamma(t)}\sigma_-\otimes\sigma_+\right]
\end{equation}
and $L(t)= \sqrt{2\gamma_0(t)}\sigma_- \otimes \mathbb{I}_2 + \sqrt{2\gamma(t)}\mathbb{I}_2  \otimes  \sigma_-$. The joint $AB$ system is initially prepared at state  $|\psi_{AB}(0)\rangle=|10\rangle$.

Consider the simple case that the coherent driving is a hard $\pi$ pulse (i.e., $u(t)=\pi\delta(t)$) under which the system $B$ is instantaneously flipped from $|0\rangle$ to $|1\rangle$, and hence $|\psi_{AB}(t)\rangle$ is flipped from $|10\rangle$ to $|11\rangle$ at $t=0$. Hence, we only need to calculate the joint system's dynamics starting from the flipped state $|\psi_{AB}(0)\rangle=|11\rangle$. It can be verified from Corollary \ref{corollary} that the wavepacket functions $\xi^{\tau_1,\cdots,\tau_\ell}$ in the steady state all vanish except when $\ell=2$, and the two-photon wavepacket can be expressed as follows:
\begin{align}\label{eq:2-photon-ex3}
\xi^{\tau_1,\tau_2}=& 2 \sqrt{{\gamma}_0(\tau_2){\gamma}(\tau_1)} e^{-{\Gamma}_A(\tau_2)-{\Gamma}_B(\tau_1)}+ 2 \sqrt{{\gamma}_0(\tau_1){\gamma}(\tau_2)} e^{-{\Gamma}_A(\tau_1)-{\Gamma}_B(\tau_2)} \nonumber\\
&+ 2 \sqrt{{\gamma}(\tau_1){\gamma}(\tau_2)} e^{-{\Gamma}_B(\tau_1)-{\Gamma}_B(\tau_2)}\left[\Xi(\tau_1)-\Xi(\tau_2)\right],
\end{align}
where
\begin{equation}\label{eq:the intermediate variable}
\Gamma_{A,B}(t)=\int_0^t\gamma_{A,B}(s)ds, \quad \Xi(t)=2\int_0^t\sqrt{\gamma_0(s)\gamma(s)}e^{\Gamma_B(s)-\Gamma_A(s)}ds.
\end{equation}

Recall that the system $A$ is chosen to generate the desired single flying-qubit wavepacket, i.e., $\xi_0^\tau=\sqrt{2{\gamma}_0(\tau)} e^{-{\Gamma}_A(\tau)}$. Let $\xi_1^\tau=\sqrt{2\gamma(\tau)}e^{-\Gamma_B(\tau)}$ be the wavepacket of a single flying-qubit spontaneously emitted from the system $B$ driven by vacuum input. We can rewrite Eq.~(\ref{eq:2-photon-ex3}) in a more compact form:
\begin{equation}\label{eq:2-photon-ex3-compact}
\xi^{\tau_1,\tau_2}=\xi_0^{\tau_1}\xi_1^{\tau_2}+\xi_0^{\tau_2}\xi_1^{\tau_1}+\xi_1^{\tau_1}\xi_1^{\tau_2}
\int_{\tau_1}^{\tau_2}\xi_0^s\xi_1^s\left[\int_{0}^s |\xi_1^\alpha|^2d\alpha\right]^{-1}.
\end{equation}
This form indicates that the incoming single photon knocks out an additional photon from the excited qubit, as is also called stimulated emission~\cite{Rephaeli2012}, and the two single photons are entangled in a complex manner.

\begin{figure}
	\centering
	\includegraphics[width=1\columnwidth]{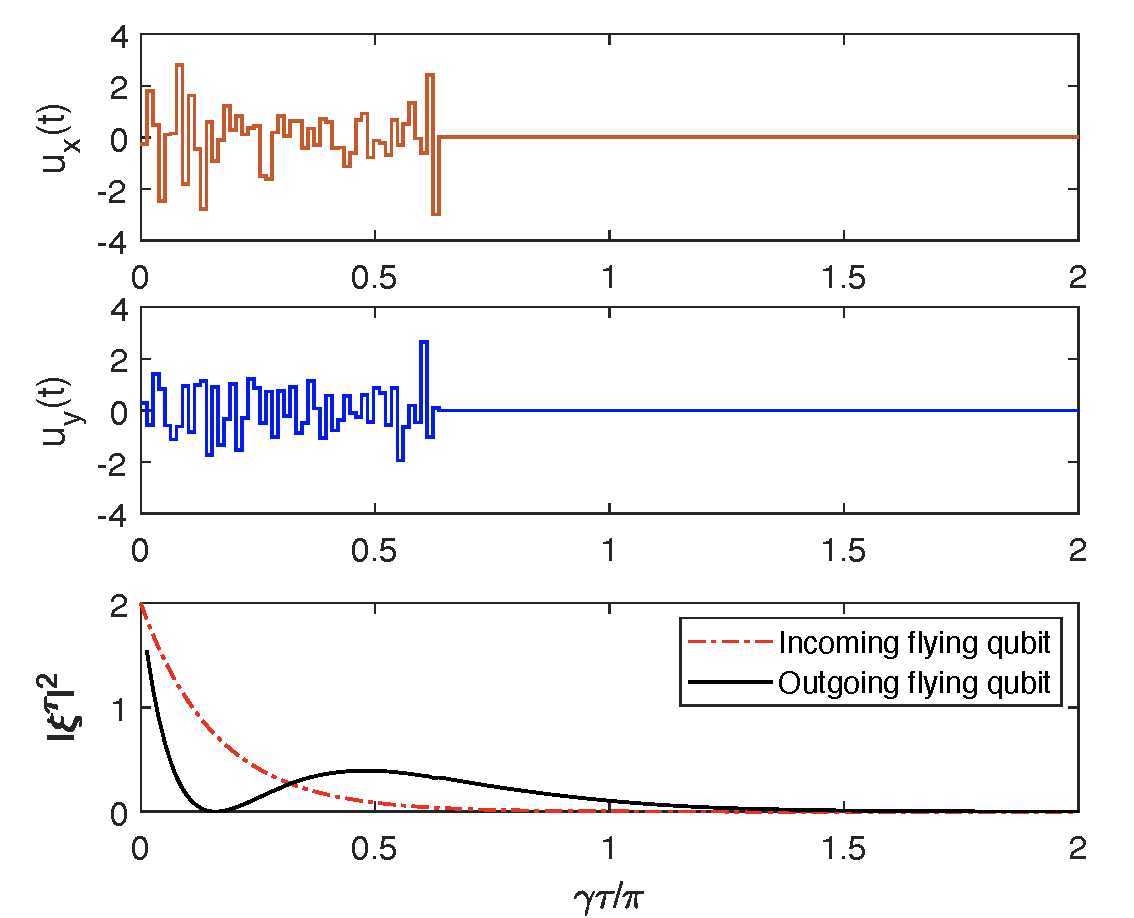}
	\caption{The transformation of a flying qubit with (a)-(b) optimized control pulses and (c) the shapes of the incoming and outgoing flying qubits.} \label{fig:6}
\end{figure}

Now we consider the soft driving control pulses, and assume that the incoming single photon has exponentially decaying waveform $\xi_0^\tau=\sqrt{2\gamma_0}e^{-\gamma_0\tau}$ (corresponding to $\epsilon(t)=0$ and $\gamma_0(t)\equiv \gamma_0$ for the auxiliary system $A$). The coherent driving control $u(t)=u_x(t)+iu_y(t)$ on the system $B$ is turned on over a fixed time interval to reshape the flying qubit, and the joint system is initially prepared at state $\ket{\Psi(0)}=\ket{10}$. We apply the genetic algorithm to optimize the waveform of $u(t)$ that maximizes the single-photon emission probability $P_1$. Numerical simulations show that the coherent control can improve the single-photon generation probability to $95.22\%$ after 200 generations, yielding the optimized control pulse displayed in Fig.~\ref{fig:6}(a). As is shown in Fig.~\ref{fig:6}(b), the shape of the outgoing single-photon component is suppressed in the beginning and then revives, showing the ability of the driving control on reshaping the incoming flying qubit.

\section{Conclusion}\label{Sec:Conclusion}
To conclude, we have developed a QSDE-based modeling method for the control of flying qubits in networked quantum information processing systems. Under the time-ordered photon-number basis, the proposed model can describe most general cases when the standing quantum system is steered by coherent or incoherent classical controls, and the quantum inputs can be either vacuum or non-vacuum. As demonstrated by examples, the states of the outgoing flying qubits can be efficiently computed by the derived low-dimensional non-unitary Schr\"{o}dinger equation and the associated quantum jumps. This QSDE-based approach can be naturally applied to the analysis flying-qubit control dynamics in quantum networks by combining the $(S,L,H)$ formulation.

The established framework also lays the foundation for practical design of flying-qubit control systems, such as the precise shaping of flying qubits, the suppression of undesired multi-photon emissions, or robust control of flying qubits. As is exemplified in the numerical example, these design problems can be formulated as properly defined optimal control problems, and efficient algorithms need to be developed. These problems have been under consideration in our future studies.

The proposed model can be naturally extended to cases when the standing quantum system has multiple levels or is coupled to multiple waveguides. However, although the analysis of flying-qubit catching and transformation problems can still be handled by introducing multiple auxiliary systems coupled to these waveguides, the resulting calculation will become extremely hard as the equivalent system defined on the tensor of multiple Hilbert spaces become large. Hence, a direct method that is free of auxiliary systems is highly demanded. This has been shown to be possible from quantum scattering theory~\cite{Hurst2018,Trivedi2018}, but the QSDE-based approach, which is potentially advantageous in the applications to complex networks, is still an open question. This will be also explored in our future works.

\bibliographystyle{plain}

\end{document}